\documentclass{article}

\usepackage[english]{babel}
\usepackage{authblk}

\usepackage[letterpaper,top=2cm,bottom=2cm,left=3cm,right=3cm,marginparwidth=1.75cm]{geometry}

\usepackage{amsmath}
\usepackage{graphicx}
\usepackage[colorlinks=true, allcolors=blue]{hyperref}
\usepackage{physics}
\usepackage{subcaption}

\begin{document}

\title{Optimal bidding in multiperiod day-ahead electricity markets assuming non-uniform uncertainty of clearing prices}
\author[1,2,*]{D\'{a}vid Csercsik}
\author[2]{Mih\'{a}ly Andr\'{a}s V\'{a}ghy}

\affil[1]{\small Institute of Economics, ELTE Centre for Economic and Regional Studies T\'oth K\'alm\'an u. 4., H-1097 Budapest, Hungary\\ \emph{csercsik.david@krtk.elte.hu}}
\affil[2]{\small P\'{a}zm\'{a}ny P\'{e}ter Catholic University, Faculty of Information Technology and Bionics, Pr\'{a}ter~u. 50/A H-1083 Budapest, Hungary\\
\emph{vaghy.mihaly.andras@itk.ppke.hu}}
\affil[*]{Corresponding author}

\maketitle

\begin{abstract}
In a recent publication, using a simple two-period model, which is already capable to capture essential non-convex multiperiod bids, Richstein et al. have shown that in the case of optimal bidding, multi-part bidding always ensures a higher expected profit for the bidder, compared to simple bidding and block-bidding. The model proposed in their analysis assumes a uniform distribution of the market-clearing prices in both periods. In this paper, we study how the conclusions of the analysis are affected, if a very simple, symmetric, stepwise-constant but non-uniform distribution is assumed in the case of the market-clearing price. We show that the results of Richstein et al. also hold in this case.
\end{abstract}

\section{Introduction}\label{sec_introduction}

Since the introduction of liberalized electricity markets \cite{cramton2017electricity}, and day-ahead power exchanges in particular \cite{Contreras2001}, profit-oriented participation in such markets is a fundamental incentive for generator companies (GenCos) \cite{Kleindorfer2001}.
As formulated in \cite{li2011modeling}, \emph{'participants} (of electricity spot markets) \emph{have to make decisions independently under complicated situations with insufficient information about their rivals and various uncertainties in the market such as load variations, competitor's behavior, and power system contingencies.'}
As a consequence, the problem of strategic or optimal bidding (i.e. decision making under the uncertainty of the market outcome) has been widely studied in the literature, using different models and various assumptions \cite{conejo2010decision,li2011modeling,prabavathi2015energy}.

\subsection{Related literature}

The reviews \cite{li2011modeling,prabavathi2015energy} classify the literature models for bidding strategy analysis and identify 3 main groups of models: (1) single GenCo optimization models, (2) game theory based models, and (3) agent-based models (the article \cite{li2011modeling} also considers a fourth group, namely 'other and hybrid models').
Single GenCo models focus on the behavior of a single supply-side player, neglecting or simplifying the strategic behavior of other market participants. One of the most important property of such models is that the market clearing price (MCP) is usually considered as an exogenous (independent) variable, in other words, the generating unit in question does not exerts market power, the bids submitted by it do not affect the result of the market outcome.
Game theory models in contrast typically assume that the market equilibrium is the direct consequence of the strategy (bid) choice of participants. Agent based models simulate the rule-based decision making process and interaction of participants (agents), usually assuming that the potentially heterogeneous agents have imperfect or local information, based on which they determine their bids.

In addition, let us note that several recent papers formulating single GenCo optimization models or agent-based models are focusing on characteristic features and the implied optimal bidding strategy of various participant types, e.g. electric vehicle aggregators \cite{vagropoulos2013optimal,vaya2014optimal} or demand side aggregators in general \cite{xu2015risk}, hydro-electric plants \cite{steeger2014optimal}, wind producers \cite{zhang2012optimal,guerrero2015optimal}, electro-voltaic producers \cite{bessa2013global} and so on.

\subsection{Motivations and aim}

A significant part of optimal bidding models in the literature consider a simplified generation cost model, in which start-up costs and variable costs of generation are not considered individually. These models usually assume that generating units are described by a single cost parameter, quantifying their generation cost per unit per bidding period, which internalizes all types of costs.
The above methodology is used, e.g. in the game theory based model \cite{hao2000study}, which approaches the problem of strategic bidding via equilibrium theory in the case of clearing price auctions, assuming a finite number of interacting participants, who do have information about each others parameters.
However, as it is described in \cite{cramton2017electricity}, the internalisation of start-up costs often has significant drawbacks.

On the other hand, while there are examples of optimal bidding models, which consider multiperiod scenarios -- see e.g. \cite{de2003simulating,de2004finding,bo2017optimal} --, these models consider the multiperiod nature of the auction rather as an iterative, repeated framework, and do not consider special non-convex orders typically used in day-ahead electricity markets, as the block order \cite{meeus2009block}, which potentially include multiple periods as well, thus define interdependencies between clearing periods \cite{madani2017revisiting}.

The recent article by Richstein et al. \cite{main_article} considers a single GenCo scenario, in a two-period market, where the participant in question has the possibility to submit either conventional single-period bids which are cleared and paid off independently for the two periods, or multi-period bids (block bid or multi-part bid), which are cleared and paid off simultaneously considering the resulting MCPs of both periods. The article takes into account the start-up and variable cost components of generation, and assumes uniform distribution of the market-clearing price in both periods considered. In this paper, closed form analytic expressions are derived and compared regarding the optimal bid values and the implied expected profit in either case of bid formats.
The main conclusion of \cite{main_article} is that the expected profit is always the highest in the case of multi-part bidding, irrespective of the cost parameters of the actuasl unit.

In the current article, we analyze how sensitive this result is with respect to the modelling assumptions, i.e. we derive a similar analysis, under a different model of the uncertainty regarding the market clearing prices, which is still simple enough to allow for analytic results.

\section{Methods}\label{methods}

We consider a simple, two-period market scenario, and a generating unit bidding in this market. Apart from assumptions regarding the distribution of the market clearing price, we use the same assumptions, notations and parameters as in \cite{main_article}.

\subsection{Cost model of the generating unit}
The generating unit in question is characterized by
a fixed, positive start-up cost $c_s$, which occurs in any case the power plant produces in one or two periods.
In addition, the plant has positive variable production cost, arising in each period in which the plant is producing, and in contrast to the start-up costs, it depends on the produced quantity (linearly): the parameter $c_v$ is the linear coefficient describing the ratio of the quantity-dependent production cost, and the produced quantity.

Following the assumptions made in \cite{main_article}, we assume that the production quantities are normalized (i.e. the unit is either producing 0 or 1 unit in each period). Thus, the $C_G$ generation cost of the unit may be described as follows
\begin{equation} \label{eq_C_G}
C_G =
\begin{cases}
   c_s + 2 c_v,\quad& \text{if the unit produces in both periods, }  \\
  c_s + c_v,\quad& \text{if the unit produces in one of the periods, } \\
  0,\quad& \text{otherwise. } \\
\end{cases}
\end{equation}

\subsection{Model of the two-period market}
We consider a simple two-period market model and assume that the market-clearing prices of the periods, denoted by $0 \leq p_1 \leq 1$ and $0 \leq p_2 \leq 1$ for period 1 and 2, respectively, are independent continuous random variables. Furthermore, we assume that the generating unit has no market power, in other words, the bids submitted by the unit in question do not affect the clearing prices.

In contrast to the original article \cite{main_article}, where the authors assumed that the distribution of market-clearing prices is uniform for each period, we assume a simple non-uniform, stepwise symmetric distribution for $p_1$ and $p_2$. This is on the one hand in accordance with the results of simulation studies described in \cite{csercsik2020two}, which shows that if all bid prices and quantities in the demand and supply side of a day-ahead market follow uniform distributions, the distribution of the market clearing price resembles to a normal distribution, i.e. values in the middle of the interval are more likely to appear.
Regarding the statistical analysis of European day-ahead market clearing prices, on may refer to \cite{huisman2013history}.

\begin{figure}[h!]
\begin{center}
  \includegraphics[width=8cm]{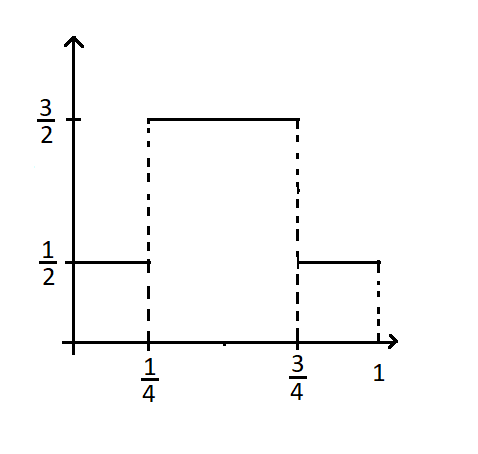}
  \caption{Stepwise distribution}
  \label{fig::stepi}
\end{center}
\end{figure}

We assume the simple stepwise distribution depicted in Figure \ref{fig::stepi} for both $p_1$ and $p_2$.
The formal equations of the probability density functions for $p_1$ and $p_2$, denoted by $\rho_{p_1}(x)$ and $\rho_{p_2}(x)$, are given as

\begin{equation}\label{eq::step_disti_1}
\rho_{p_1}(x) = \rho_{p_2}(x)=
\begin{cases}
  \frac{1}{2},\quad& 0 \leq x < \frac{1}{4}, \\
  \frac{3}{2},\quad& \frac{1}{4} \leq x < \frac{3}{4}, \\
  \frac{1}{2},\quad& \frac{3}{4} \leq x < 1, \\
  0, &  \text{otherwise.}
\end{cases}
\end{equation}

Let us note that the parameters of the proposed stepwise distribution (i.e. breakpoints and step-values) may be chosen arbitrarily. The results to be presented may be adjusted more or less straightforwardly to similar stepwise distributions with other parameters as well. By altering the assumption regarding the distribution of $p_1$ and $p_2$, the model may be considered more realistic, but in the same time, as we will see, it remains still simple enough to obtain analytical results regarding the optimal bid values.

\subsection{Simple bidding}
In the simple bid format, the generating unit bids for each period separately with bids characterized by the prices $b_1$ and $b_2$. If the realised market clearing prices $p_1$ and $p_2$ are equal or greater than the respective bid prices $p_1$ and $p_2$, both bids are accepted and the unit needs to produce in both periods with the associated costs. In this case, the unit is paid off according to the market-clearing prices $p_1$ and $p_2$, i.e. the income of the unit ($I$) equals $p_1+p_2$. If the market-clearing price exceeds the bid price in only one of the periods, the respective can be individually accepted, while the other bid is rejected, so that a unit may produce in one period, but not in the other. Let $\pi_{S}$ denote the profit for an actor in the case of simple bidding. Considering the cost model described in Eq. \eqref{eq_C_G}, it can be calculated as
\begin{equation} \label{eq:simple:pi}
\pi_{S}(b_1,b_2, p_1, p_2) = I-C_G=
\begin{cases}
  p_1 + p_2 - c_s - 2 c_v,\quad& p_1 \geq b_1,~p_2 \geq b_2, \\
  p_1 - c_s - c_v,\quad& p_1 \geq b_1,~p_2 < b_2, \\
  p_2 - c_s - c_v,\quad& p_1 < b_1,~p_2 \geq b_2, \\
  0,\quad& \text{otherwise.} \\
\end{cases}
\end{equation}
Following \cite{main_article}, we assume $b_1=b_2=b$ and define $\pi_S(b,p_1,p_2)=\pi_S(b,b,p_1,p_2)$. The expected profit $E[\pi_{S}(b)]$, as a function of $b$, can be given as
\begin{equation}\label{eq:simple:Epi}
\begin{aligned}
E[\pi_{S}(b)]&= \int_0^{1} \int_0^{1} \pi_{S}(b,p_1,p_2) \rho_{p_1}(x) \rho_{p_2}(y)\dd{x}\dd{y}\\
&= \int_b^{1} \int_b^{1} (p_1+p_2-c_s-2c_v)~ \rho_{p_1}(x) \rho_{p_2}(y)\dd{x}\dd{y}\\
&+ \int_0^{b} \int_b^{1}  (p_1-c_s-c_v)~ \rho_{p_1}(x) \rho_{p_2}(y)\dd{x}\dd{y}\\
&+ \int_b^{1} \int_0^{b} (p_2-c_s-c_v)~\rho_{p_1}(x) \rho_{p_2}(y)\dd{x}\dd{y}.
\end{aligned}
\end{equation}

\subsection{Block bidding}
In the case of block bidding, a single bid including both periods is submitted, which is either fully accepted or fully rejected. Hence, if the block bid is accepted, the unit is producing in both periods.
The profit for block bidding is denoted by $\pi_{B}$ and may be calculated as follows
\begin{equation} \label{eq:block:pi}
\pi_{B}(b, p_1, p_2) =
\begin{cases}
  p_1 + p_2 - c_s - 2 c_v,\quad& p_1 + p_2 \geq b, \\
  0,\quad& \text{otherwise. } \\
\end{cases}
\end{equation}
The expected profit $E[\pi_{B}(b)]$ can be computed as
\begin{equation}\label{eq:block:Epi}
\begin{aligned}
E[\pi_{B}(b)]&=\int_0^{1} \int_0^{1} \pi_{B}(b,p_1,p_2) \rho_{p_1}(x) \rho_{p_2}(y)\dd{x}\dd{y}\\
&=\int_0^b\int_{b-y}^1(p_1+p_2-c_s-2c_v)\rho_{p_1}(x)\rho_{p_2}(y)\dd{x}\dd{y}\\
&+\int_b^1\int_0^1(p_1+p_2-c_s-2c_v)\rho_{p_1}(x)\rho_{p_2}(y)\dd{x}\dd{y}.
\end{aligned}
\end{equation}

\subsection{Multi-part bidding} \label{multipartmethod}
Multi-part bidding allows the participant to explicitly bid the values of the start-up cost $b_s$ and variable cost $b_v$ that are valid for
both periods. Dependent on market prices the bid will be accepted in one, two, or no periods.
The profit for multi-part bidding is denoted by $\pi_{M}$ and can be given as
\begin{equation} \label{eq:multi:pi}
\pi_{M}(b_s, b_v, p_1, p_2) =
\begin{cases}
  p_1 - c_s - c_v,\quad& p_1 \geq b_s + b_v,~p_2 < b_v, \\
  p_2 - c_s - c_v,\quad& p_1<b_v,~p_2 \geq b_s + b_v,\\
  p_1 + p_2 - c_s - 2c_v,\quad& p_1,p_2\ge b_v,~p_1 + p_2 \geq b_s + 2b_v,\\
  0,\quad& \text{otherwise. } \\
\end{cases}
\end{equation}
In this case the expected profit can be described as
\begin{equation}\label{eq:multi:Epi}
E[\pi_{M}(b_s, b_v)] = \int_0^{1} \int_0^{1} \pi_{M}(b_s, b_v ,p_1,p_2) \rho_{p_1}(x) \rho_{p_2}(y)\dd{x}\dd{y}. \\
\end{equation}
We have to compute the integrals for the following 21 cases:
\begin{enumerate}
    \item $0\le b_v<\frac{1}{4}$
    \begin{enumerate}
        \item $0\le b_s\le\frac{1}{4}-b_v$
        \item $\frac{1}{4}-b_v\le b_s<\frac{1}{2}-2b_v$
        \item $\frac{1}{2}-2b_v\le b_s<\frac{3}{4}-b_v$
        \item $\frac{3}{4}-b_v\le b_s<1-2b_v$
        \item $1-2b_v\le b_s<1-b_v$
        \item $1-b_v\le b_s<\frac{5}{4}-2b_v$
        \item $\frac{5}{4}-2b_v\le b_s<\frac{3}{2}-2b_v$
        \item $\frac{3}{2}-2b_v\le b_s<\frac{7}{4}-2b_v$
        \item $\frac{7}{4}-2b_v\le b_s<2-2b_v$
    \end{enumerate}
    \item $\frac{1}{4}\le b_v<\frac{1}{2}$
    \begin{enumerate}
        \item $0\le b_s<\frac{3}{4}-b_v$
        \item $\frac{3}{4}-b_v\le b_s<1-b_v$
        \item $1-b_v\le b_s<\frac{3}{2}-2b_v$
        \item $\frac{3}{2}-2b_v\le b_s<\frac{7}{4}-2b_v$
        \item $\frac{7}{4}-2b_v\le b_s<2-2b_v$
    \end{enumerate}
    \item $\frac{1}{2}\le b_v<\frac{3}{4}$
    \begin{enumerate}
        \item $0\le b_s<\frac{3}{4}-b_v$
        \item $\frac{3}{4}-b_v\le b_s<\frac{3}{2}-2b_v$
        \item $\frac{3}{2}-2b_v\le b_s<1-b_v$
        \item $1-b_v\le b_s<\frac{7}{4}-2b_v$
        \item $\frac{7}{4}-2b_v\le b_s<2-2b_v$
    \end{enumerate}
    \item $\frac{3}{4}\le b_s<1$
    \begin{enumerate}
        \item $0\le b_s<1-b_v$
        \item $1-b_v\le b_s<2-2b_v$
    \end{enumerate}
\end{enumerate}

\section{Results}

\section{Results}

\subsection{Simple bidding}\label{simpleanalitical}
In this subsection, we derive the optimal bidding strategy in the case of simple bidding, assuming the stepwise distribution of $p_1$ and $p_2$ as described in Eq. \eqref{eq::step_disti_1}.


Computing the integrals in Eq. \eqref{eq:simple:pi} yields
\begin{equation}\label{eq:simple:Ecomp}
E[\pi_{S}(b)] =
\begin{cases}
 \displaystyle\frac{1}{4}\qty\big(-2b^2+b^2c_s+4bc_v-8c_v-4c_s+4),\quad& 0 \leq b < \frac{1}{4},\\[1em]
 \begin{aligned}[c]
 \frac{1}{16}\Big(-24b^2+36b^2c_s+48bc_v\\
 -12bc_s-40c_v-15c_s+17\Big),
 \end{aligned}
 \quad& \frac{1}{4} \leq b < \frac{3}{4}, \\[1em]
 \displaystyle\frac{1}{4}\qty\big(-2b^2+b^2c_s+4bc_v+2bc_s-4c_v-3c_s+2), & \frac{3}{4} \leq b \leq 1.
\end{cases}
\end{equation}
The algebraic details of the calculations are described in Appendix A. To determine the optimal bid, we need to find the maximum of Eq. \eqref{eq:simple:Ecomp}. Since the expected profit is given as a piecewise function, the set of critical points consists of not just the extrema of the three cases, but also of the points $b=0$, $b=\frac{1}{4}$, $b=\frac{3}{4}$ and $b=1$.

The derivatives are given as
\begin{equation}
\dv{}{b}E[\pi_{S}(b)] =
\begin{cases}
 \displaystyle\frac{1}{2} b (c_s-2) + c_v,\quad& 0 \leq b < \frac{1}{4},\\[1em]
 \displaystyle\frac{3}{4}\qty\big(b (6 c_s - 4) - c_s + 4 c_v),\quad& \frac{1}{4} \leq b < \frac{3}{4}, \\[1em]
 \displaystyle\frac{1}{2} (b (-2 + c_s) + c_s + 2 c_v), & \frac{3}{4} \leq b \leq 1.
\end{cases}
\end{equation}

\begin{figure}[h!]
\centering
\includegraphics[width=0.7\textwidth]{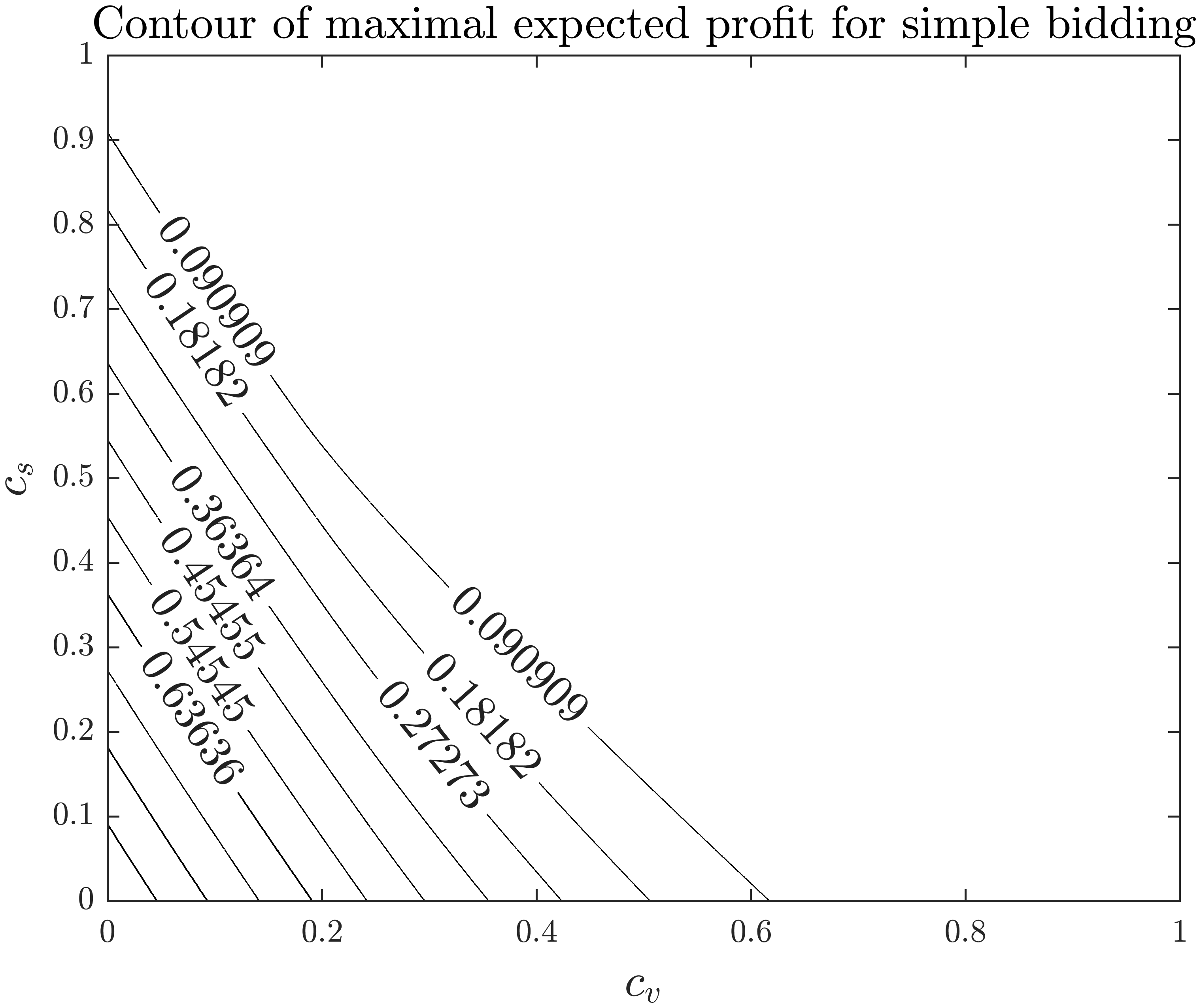}
\caption{Contour plot of the maximal expected profit for simple bidding.}
\label{fig:simple:optimal}
\end{figure}

Then the set of critical points is
\begin{equation}
\qty\bigg{0,\frac{1}{4}, \frac{3}{4}, 1, \frac{-2c_v}{c_s-2},\frac{-c_s+4c_v}{6c_s-4},\frac{-c_s-2c_v}{c_s-2}},
\end{equation}
and thus the optimal bidding strategy is
\begin{equation}
b_S^* = \max\qty\Bigg(E[\pi_{S}(b)]\bigg| b \in \qty\bigg{0,\frac{1}{4}, \frac{3}{4}, 1, \frac{-2c_v}{c_s-2},\frac{-c_s+4c_v}{6c_s-4},\frac{-c_s-2c_v}{c_s-2}}).
\end{equation}
Note, that for example in case A it might happen that the optimal $b$ value of $-\frac{2c_v}{c_s-2}$ is not in the corresponding interval $0\le b<\frac{1}{4}$. In this case that part of $E[\pi_{S}]$ reaches its maximal value at one of the endpoints of the underlying interval and the function value at $-\frac{2c_v}{c_s-2}$ need not to be checked. Figure \ref{fig:simple:optimal} shows the contour plot of the maximal expected profit as a function of $c_s$ and $c_v$.

\subsection{Block bidding}
\begin{figure}[h!]
\centering
\includegraphics[width=0.7\textwidth]{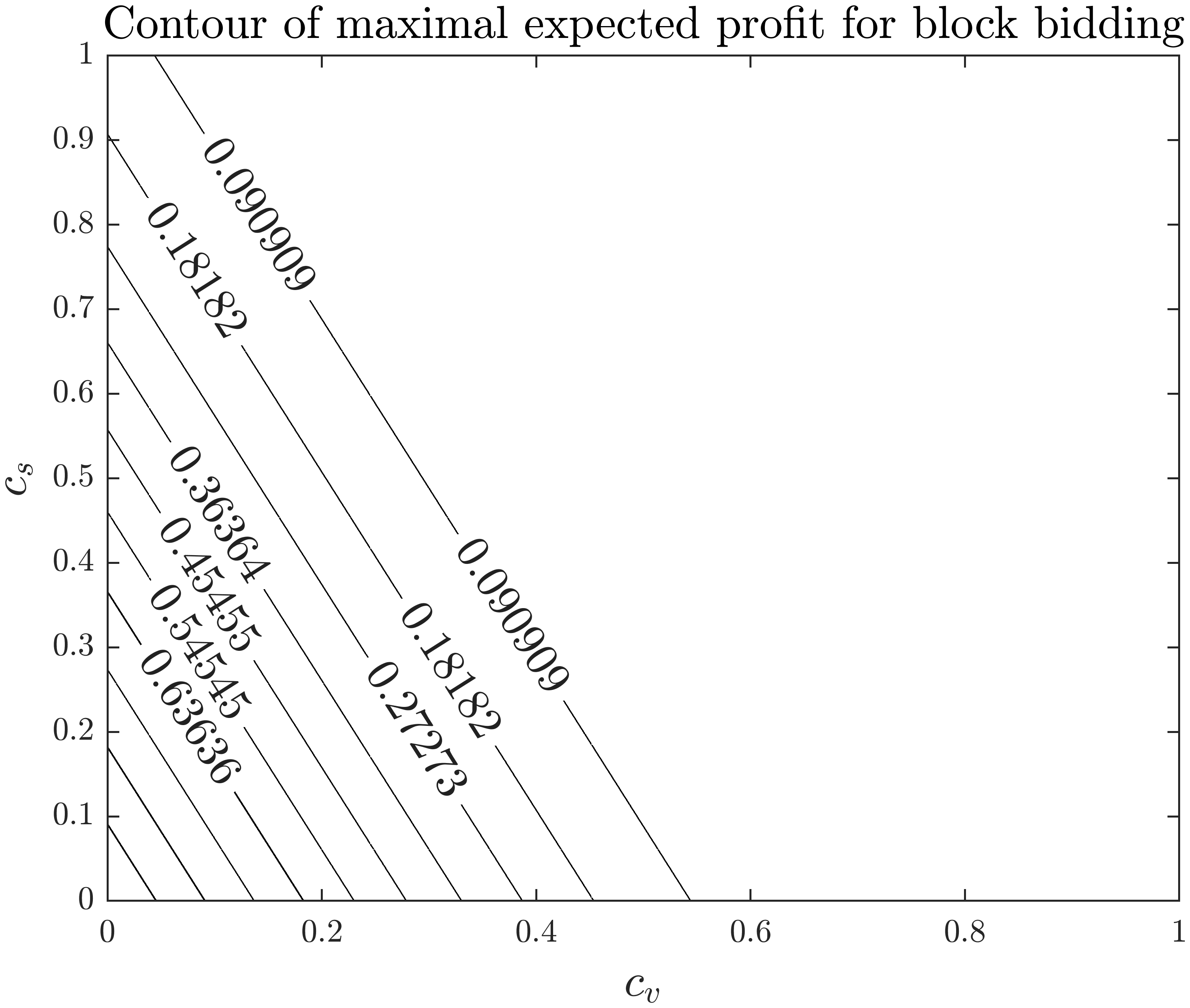}
\caption{Contour plot of the maximal expected profit for block bidding.}
\label{fig:block:optimal}
\end{figure}

Computing the integrals in Eq. \eqref{eq:block:Epi} yields
\begin{equation}\label{eq:block:Ecomp}
E[\pi_{B}(b)] =
\begin{cases}
    \displaystyle-\frac{b^3}{12}+\frac{b^2(c_s+2c_v)}{8}-2c_v-c_s+1,\quad& 0 \leq b < \frac{1}{4},\\[1em]
    \begin{aligned}[c]
        &-\frac{5b^3}{12}+\frac{b^2(5c_s+10c_v+1)}{8}\\
        &-\frac{b(c_s+2c_v)}{4}-\frac{31c_s+62c_v}{32}+\frac{383}{384},
    \end{aligned}
    \quad& \frac{1}{4} \leq b < \frac{1}{2}, \\[2em]
    \begin{aligned}[c]
        &-\frac{3b^3}{4}+\frac{b^2(9c_s+18c_v+4)}{8}\\
        &-\frac{b(3c_s+6c_v)}{4}-\frac{27c_s+54c_v}{32}+\frac{125}{128},
    \end{aligned}
    \quad& \frac{1}{2} \leq b < \frac{3}{4}, \\[2em]
    \displaystyle-\frac{5b^3}{12}+\frac{b^2(5c_s+10c_v)}{8}-\frac{9c_s+18c_v}{8}+\frac{67}{64},\quad& \frac{3}{4} \leq b < 1, \\[1em]
    \begin{aligned}[c]
        &\frac{5b^3}{12}-\frac{b^2(5c_s+10c_v+10)}{8}\\
        &+\frac{b(5c_s+10c_v)}{2}-\frac{19c_s+38c_v}{8}+\frac{281}{192},
    \end{aligned}
    \quad& 1 \leq b < \frac{5}{4}, \\[2em]
    \begin{aligned}[c]
        &\frac{3b^3}{4}-\frac{b^2(9c_s+18c_v+15)}{8}\\
        &+\frac{b(15c_s+30c_v)}{4}-\frac{101c_s+202c_v}{32}+\frac{229}{128},
    \end{aligned}
    \quad& \frac{5}{4} \leq b < \frac{3}{2}, \\[2em]
    \begin{aligned}[c]
        &\frac{5b^3}{12}-\frac{b^2(5c_s+10c_v+9)}{8}\\
        &+\frac{b(9c_s+18c_v)}{4}-\frac{65c_s+130c_v}{32}+\frac{157}{128},
    \end{aligned}
    \quad& \frac{3}{2} \leq b < \frac{7}{4}, \\[2em]
    \displaystyle\frac{b^3}{12}-\frac{b^2(c_s+2c_v+2)}{8}+\frac{b(c_s+2c_v)}{2}-\frac{c_s+2c_v}{2}+\frac{1}{3},\quad& \frac{7}{4} \leq b < 2.
\end{cases}
\end{equation}
Figure \ref{fig:block:optimal} shows the contour plot of the expected profit as a function of $c_s$ and $c_v$.

The derivatives are given as
\begin{equation}
\dv{}{b}E[\pi_{B}(b)] =
\begin{cases}
    \displaystyle-\frac{b^2}{4}+\frac{b(c_s+2c_v}{2},\quad& 0 \leq b < \frac{1}{4},\\[1em]
    \displaystyle-\frac{5b^2}{4}+\frac{b(5c_s+10c_v+1)}{4}-\frac{c_s+2c_v}{4}\quad& \frac{1}{4} \leq b < \frac{1}{2}, \\[1em]
    \displaystyle-\frac{9b^3}{4}+\frac{b(9c_s+9c_v+3}{4}-\frac{3c_s+6c_v}{4}\quad& \frac{1}{2} \leq b < \frac{3}{4}, \\[1em]
    \displaystyle-\frac{5b^2}{4}+\frac{b(5c_s+10c_v)}{4},\quad& \frac{3}{4} \leq b < 1, \\[1em]
    \displaystyle\frac{5b^2}{4}-\frac{b(5c_s+10c_v+10)}{4}+\frac{5c_s+10c_v}{2}\quad& 1 \leq b < \frac{5}{4}, \\[1em]
    \displaystyle\frac{9b^2}{4}-\frac{b(9c_s+18c_v+15)}{4}+\frac{15c_s+30c_v}{4}\quad& \frac{5}{4} \leq b < \frac{3}{2}, \\[1em]
    \displaystyle\frac{5b^2}{4}-\frac{b(5c_s+10c_v+9}{4}+\frac{9c_s+18c_v}{4}\quad& \frac{3}{2} \leq b < \frac{7}{4}, \\[1em]
    \displaystyle\frac{b^2}{4}-\frac{b(c_s+2c_v+2)}{4}+\frac{c_s+2c_v}{2},\quad& \frac{7}{4} \leq b < 2.
\end{cases}
\end{equation}
Then the set of critical points is
\begin{equation}
\qty\bigg{0,\frac{1}{4}, \frac{1}{2},\frac{3}{4}, 1, \frac{5}{4},\frac{3}{2},\frac{7}{4},2,c_s+2c_v},
\end{equation}
and thus the optimal bidding strategy is
\begin{equation}
b_B^* = \max\qty\Bigg(E[\pi_{B}(b)]\bigg| b \in \qty\bigg{0,\frac{1}{4}, \frac{1}{2},\frac{3}{4}, 1, \frac{5}{4},\frac{3}{2},\frac{7}{4},2,c_s+2c_v}).
\end{equation}

\subsection{Multi-part bidding}

\begin{figure}[h!]
\centering
\includegraphics[width=0.7\textwidth]{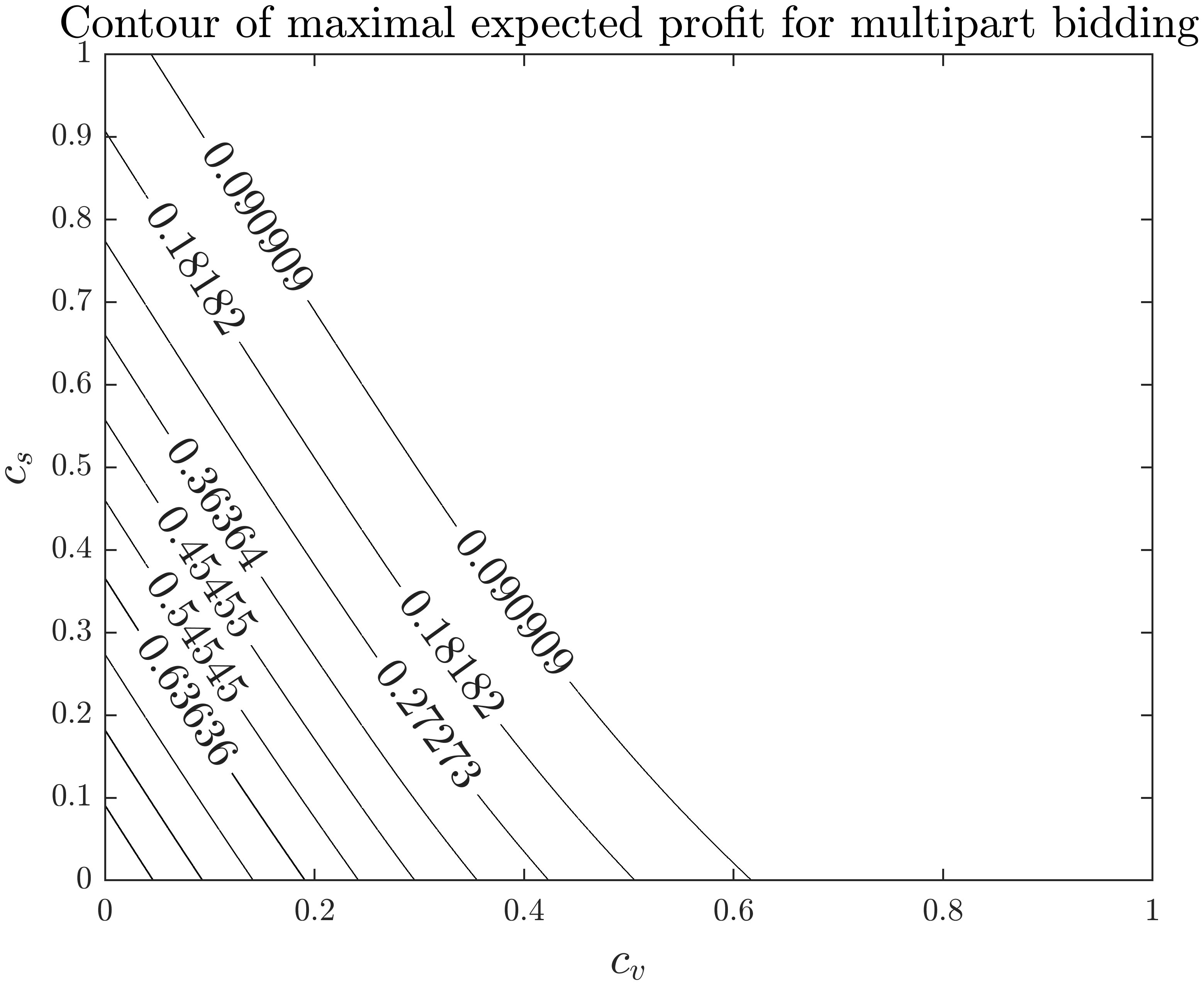}
\caption{Contour plot of the maximal expected profit for multi-part bidding.}
\label{fig:multi:optimal}
\end{figure}

Figure \ref{fig:multi:optimal} shows the contour plot of the expected profit as a function of $c_s$ and $c_v$.

\subsection{Comparison}
\begin{figure}[h!]
\centering
\includegraphics[width=0.7\textwidth]{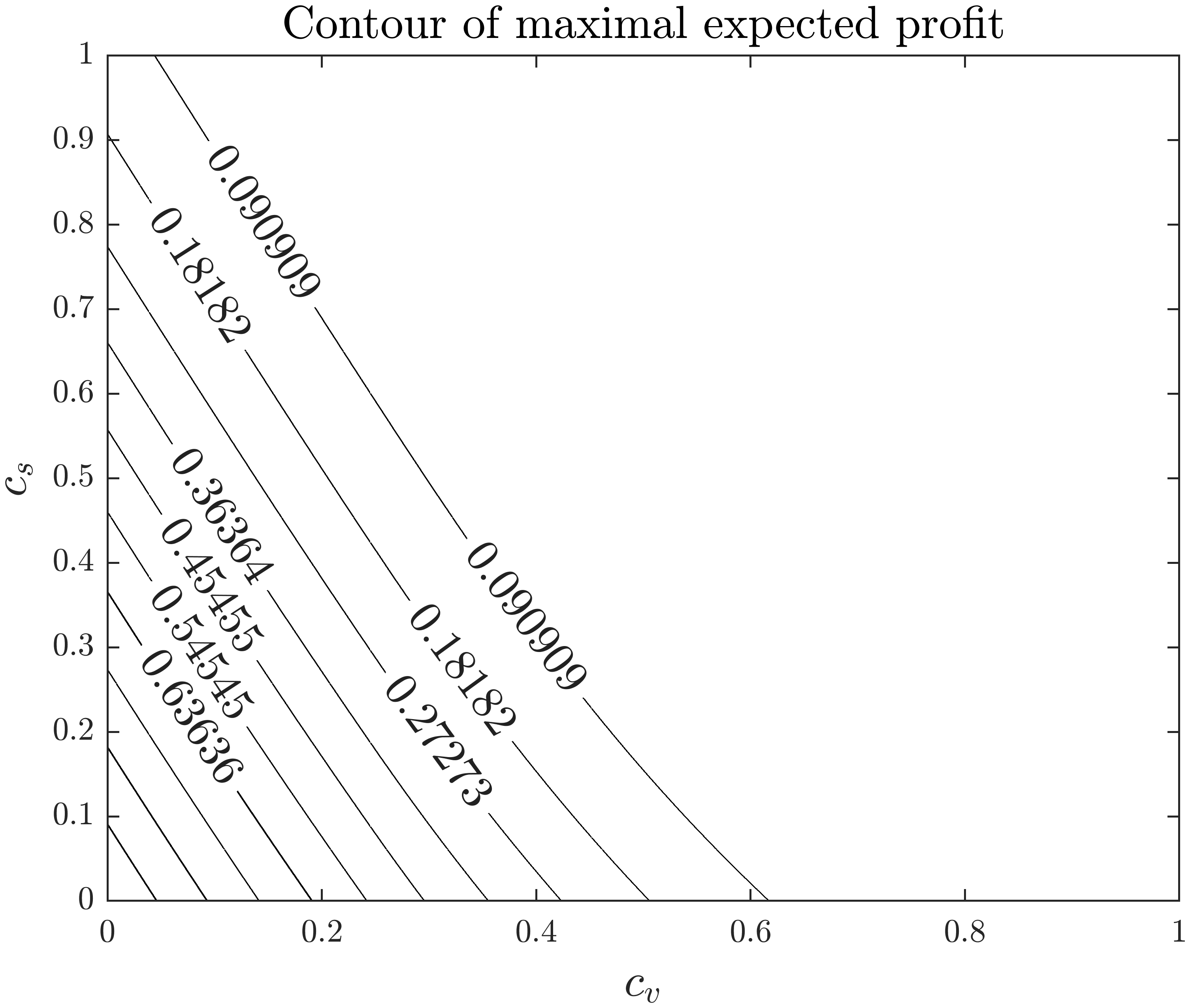}
\caption{Contour plot of the maximal expected profit, considering all possible bidding formats.}
\label{fig:combined:optimal}
\end{figure}

Figure \ref{fig:combined:optimal} depicts the contour of maximal expected profits, as a function of $c_s$ and $c_v$, considering all possible bidding formats. As one may notice, this figure aligns with Fig. \ref{fig:multi:optimal}.

\begin{figure}[h!]
\centering
\begin{subfigure}[b]{0.49\textwidth}
    \centering
    \includegraphics[width=\textwidth]{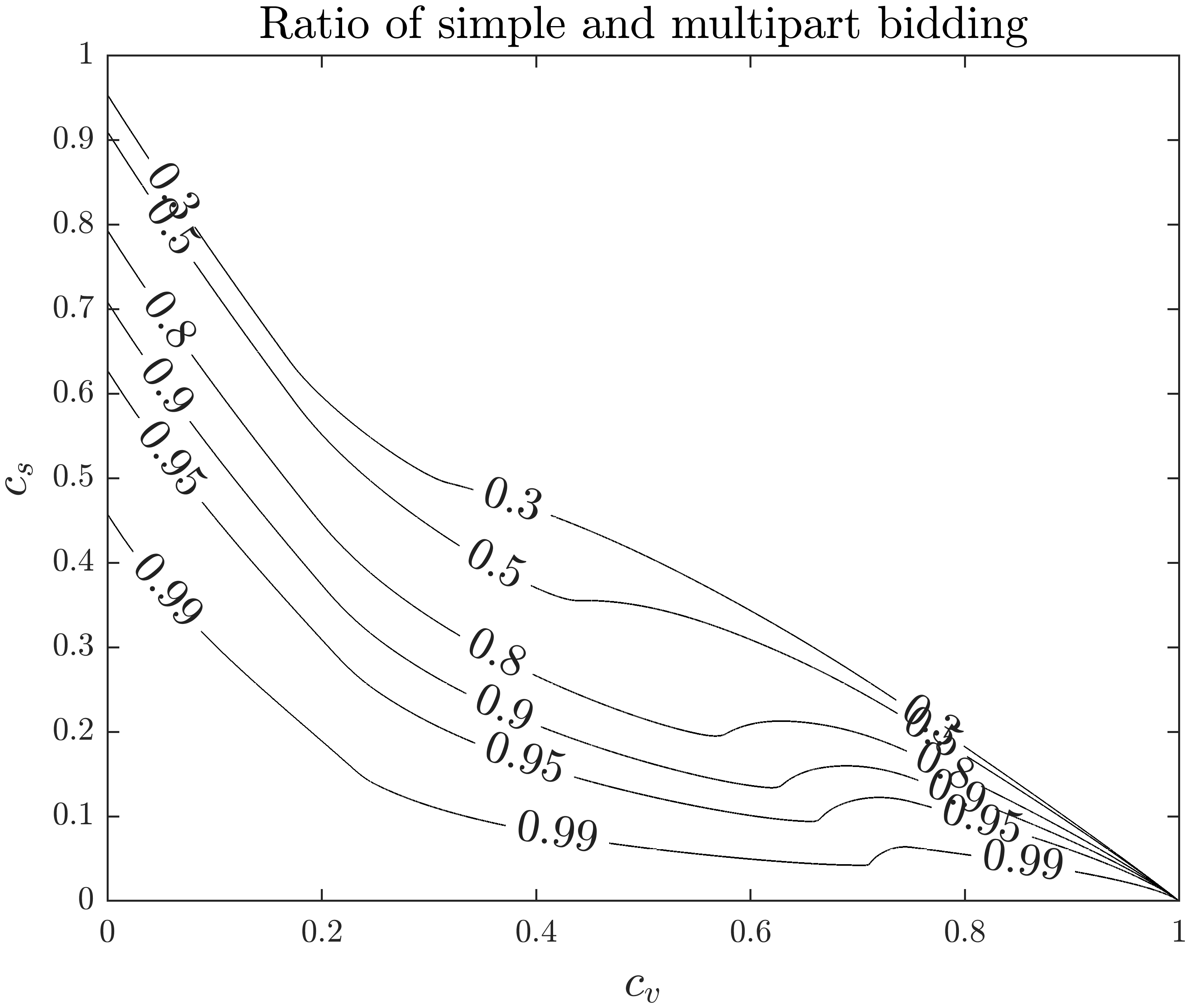}
 \end{subfigure}
 \hfill
 \begin{subfigure}[b]{0.49\textwidth}
    \centering
    \includegraphics[width=\textwidth]{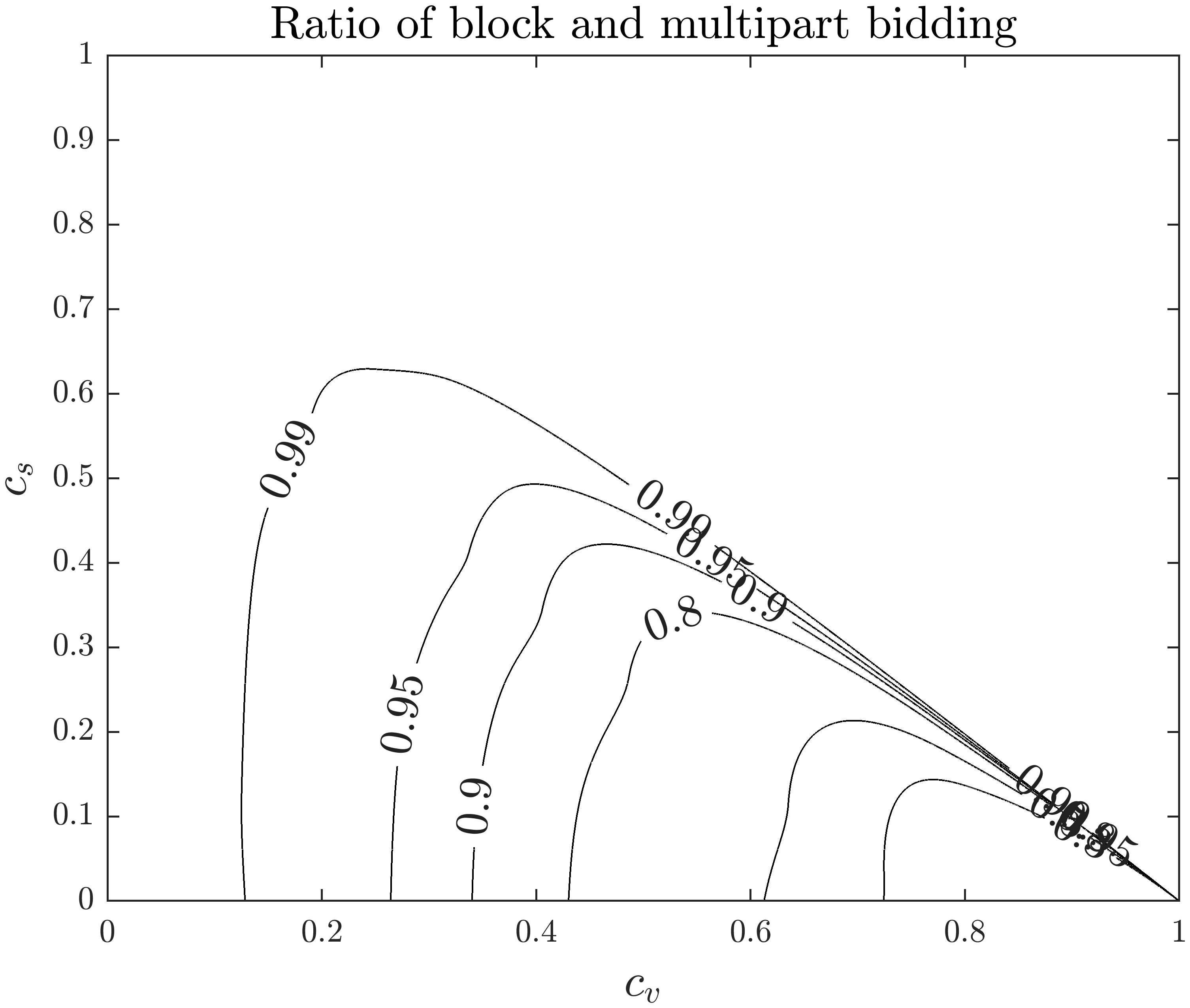}
 \end{subfigure}
 \caption{Contour plots of ratios of simple vs. multipart bidding and block vs multipart bidding.}
\label{fig:ratios}
\end{figure}

\section{Conclusions}

The article \cite{main_article} has shown that multipart bidding results in higher expected profits compared to simple and block bidding, independent of the parameters $c_s$ and $c_v$, when assuming uniform distribution of the market-clearing prices. We repeated the analysis for a simple case of non-uniform market-clearing prices, where the density function is piecewise constant and symmetric. Although analytical results for the relation of the bidding formats are not presented in the current article, the numerical results derived are in line with the previous results of \cite{main_article} in the sense that multipart bidding proves to be optimal also in this case of the clearing price distribution. Although an analytical proof is nontrivial even in this case, further work is required to decide whether the optimality of multipart bidding is valid independent of the distribution of the market-clearing price.

\section{Acknowledgements}
This work has been supported by the Fund FK 137608 of the Hungarian National Research, Development and Innovation Office.

\section{Acknowledgements}
This work has been supported by the Fund FK 137608 of the Hungarian National Research, Development and Innovation Office.

\bibliographystyle{plain}
\bibliography{ref_plus_old.bib}

\section*{Appendix A}\

\subsection{Case A}  \label{Append Case A}
\begin{small}
In this case we suppose that $0 \leq b < \frac{1}{4}$, thus \eqref{eq:simple:Epi} may be written as
{\small
\begin{align*}
& \int_b^{1} \left(
    \int_b^{\frac{1}{4}}
        \frac{1}{2}(x+y-c_s-2c_v) dx
    + \int_{\frac{1}{4}}^{\frac{3}{4}}
        \frac{3}{2}(x+y-c_s-2c_v) dx
\right)\rho_{p_2}(y)  ~ dy  + \\
& \int_b^{1} \left(
    \int_{\frac{3}{4}}^{1}
        \frac{1}{2}(x+y-c_s-2c_v) dx
\right)\rho_{p_2}(y)  ~ dy  + \\
&  \int_0^{b} \left(
    \int_b^{\frac{1}{4}}
        \frac{1}{2}(x-c_s-c_v) dx
    + \int_{\frac{1}{4}}^{\frac{3}{4}}
        \frac{3}{2}(x-c_s-c_v) dx
    + \int_{\frac{3}{4}}^{1}
        \frac{1}{2}(x-c_s-c_v) dx
\right)\rho_{p_2}(y) ~ dy  + \\
&  \int_b^{1} \left(
    \int_0^{b}
        \frac{1}{2}(y-c_s-c_v) dx
\right) \rho_{p_2}(y) ~ dy =
\end{align*}

\begin{align*}
& \int_b^{\frac{1}{4}}
    \int_b^{\frac{1}{4}}
        \frac{1}{2}\frac{1}{2}(x+y-c_s-2c_v) dx dy
  +\int_b^{\frac{1}{4}}
     \int_{\frac{1}{4}}^{\frac{3}{4}}
        \frac{1}{2}\frac{3}{2}(x+y-c_s-2c_v) dx dy + \\
&   +\int_b^{\frac{1}{4}}
    \int_{\frac{3}{4}}^{1}
        \frac{1}{2}\frac{1}{2}(x+y-c_s-2c_v) dx dy
  \int_{\frac{1}{4}}^{\frac{3}{4}}
    \int_b^{\frac{1}{4}}
        \frac{3}{2}\frac{1}{2}(x+y-c_s-2c_v) dx dy +\\
&  +\int_{\frac{1}{4}}^{\frac{3}{4}}
     \int_{\frac{1}{4}}^{\frac{3}{4}}
        \frac{3}{2}\frac{3}{2}(x+y-c_s-2c_v) dx dy
   +\int_{\frac{1}{4}}^{\frac{3}{4}}
    \int_{\frac{3}{4}}^{1}
        \frac{3}{2}\frac{1}{2}(x+y-c_s-2c_v) dx dy   +\\
& \int_{\frac{3}{4}}^{1}
    \int_b^{\frac{1}{4}}
        \frac{1}{2}\frac{1}{2}(x+y-c_s-2c_v) dx dy
  +\int_{\frac{3}{4}}^{1}
     \int_{\frac{1}{4}}^{\frac{3}{4}}
        \frac{1}{2}\frac{3}{2}(x+y-c_s-2c_v) dx dy   +\\
&   +\int_{\frac{3}{4}}^{1}
    \int_{\frac{3}{4}}^{1}
        \frac{1}{2}\frac{1}{2}(x+y-c_s-2c_v) dx dy
  \int_0^{b}
    \int_b^{\frac{1}{4}}
        \frac{1}{2}\frac{1}{2}(x-c_s-c_v) dx dy   +\\
& + \int_0^{b}
     \int_{\frac{1}{4}}^{\frac{3}{4}}
       \frac{1}{2} \frac{3}{2}(x-c_s-c_v) dx dy
 + \int_0^{b}
    \int_{\frac{3}{4}}^{1}
        \frac{1}{2}\frac{1}{2}(x-c_s-c_v) dx  dy   +\\
&  \int_b^{\frac{1}{4}}
    \int_0^{b}
        \frac{1}{2}\frac{1}{2}(y-c_s-c_v) dx  dy
+ \int_{\frac{1}{4}}^{\frac{3}{4}}
    \int_0^{b}
        \frac{3}{2}\frac{1}{2}(y-c_s-c_v) dx  dy   +\\
&+ \int_{\frac{3}{4}}^{1}
    \int_0^{b}
        \frac{1}{2}\frac{1}{2}(y-c_s-c_v) dx  dy=
\end{align*}

Wolram Alpha \cite{Wolfram} has been used for algebraic simplification.

\begin{align*}
&\frac{1}{256}(1 - 4 b)^2 (4 b - 4 c_s - 8 c_v + 1) - \frac{3}{256} (4 b - 1) (4 b - 8 c_s - 16 c_v + 5) \\
&- \frac{1}{128} (4 b - 1) (b - 2 c_s - 4 c_v + 2)- \frac{3}{256} (4 b - 1) (4 b - 8 c_s - 16 c_v + 5) \\
& - \frac{9}{16} (c_s + 2 c_v - 1) - \frac{1}{256} (8 c_s + 16 c_v - 11)- \frac{1}{128} (4 b - 1) (b - 2 c_s - 4 c_v + 2)\\
& - \frac{3}{256} (8 c_s + 16 c_v - 11) + \frac{1}{256} (-4 c_s - 8 c_v + 7) - \frac{1}{128} b (4 b - 1) (4 b - 8 c_s - 8 c_v + 1) \\
& - \frac{3}{16} b (2 c_s + 2 c_v - 1) - \frac{1}{128} b (8 c_s + 8 c_v - 7)- \frac{1}{128} b (4 b - 1) (4 b - 8 c_s - 8 c_v + 1) \\
& - \frac{3}{16} b (2 c_s + 2 c_v - 1) - \frac{1}{128} b (8 c_s + 8 c_v - 7)=
\end{align*}






$$
=\frac{-2b^2 +b^2c_s+4bc_v-4c_s-8c_v + 4}{4}
$$
}
\end{small}
\subsection{Case B} \label{Append Case B}
\begin{small}
In this case we suppose that $\frac{1}{4} \leq b < \frac{3}{4}$, thus Eq. \eqref{eq:simple:Epi} may be written as

\begin{align*}
&\int_b^{1}    \left(
    \int_{b}^{\frac{3}{4}}
        \frac{3}{2}(x+y-c_s-2c_v) dx
    + \int_{\frac{3}{4}}^{1}
        \frac{1}{2}(x+y-c_s-2c_v) dx
\right) \rho_{p_2}(y)  ~ dy + \\
&  \int_0^{b} \left(
    \int_{b}^{\frac{3}{4}}
        \frac{3}{2}(x-c_s-c_v) dx
    + \int_{\frac{3}{4}}^{1}
        \frac{1}{2}(x-c_s-c_v) dx
\right) \rho_{p_2}(y)  ~ dy + \\
&  \int_b^{1} \left(
    \int_0^{\frac{1}{4}}
        \frac{1}{2}(y-c_s-c_v) dx
    + \int_{\frac{1}{4}}^b
        \frac{3}{2}(y-c_s-c_v) dx
\right) \rho_{p_2}(y) ~ dy
\end{align*}

Which is equal to:
$$\frac{-24b^2+36b^2c_s+48bc_v-12bc_s-40c_v-15c_s+17}{16}$$
\end{small}
\subsection{Case C}  \label{Append Case C}
\begin{small}
In this case we suppose that $\frac{3}{4} \leq b \leq 1$, thus Eq. \eqref{eq:simple:Epi} may be written as
\begin{align*}
&\int_b^{1} \left(
    \int_{b}^{1}
        \frac{1}{2}(x+y-c_s-2c_v) dx
\right) \rho_{p_2}(y)  ~ dy  +\\
&  \int_0^{b} \left(
    \int_{b}^{1}
        \frac{1}{2}(x-c_s-c_v) dx
\right) \rho_{p_2}(y)  ~ dy +\\
&  \int_b^{1} \left(
    \int_0^{\frac{1}{4}}
        \frac{1}{2}(y-c_s-c_v) dx
    + \int_{\frac{1}{4}}^{\frac{3}{4}}
        \frac{3}{2}(y-c_s-c_v) dx +
    \int_{\frac{3}{4}}^{b}
        \frac{1}{2}(y-c_s-c_v) dx
\right) \rho_{p_2}(y) ~ dy
\end{align*}

Which is equal to:
$$\frac{-2b^2+b^2c_s+4bc_v+2bc_s-4c_v-3c_s+2}{4}$$
\end{small}

\subsection{Algebraic details in the case of multi-part bidding}
\begin{small}
\begin{align*}
        E[\pi_M(b_s,b_v)]&=-\frac{{b_s}^3}{12}-\frac{{b_s}^2b_v}{2}+\frac{{b_s}^2c_s}{8}+\frac{{b_s}^2c_v}{4}-\frac{b_s{b_v}^2}{2}+\frac{b_sb_vc_s}{2}+\frac{b_sb_vc_v}{2}\\
        &+\frac{{b_v}^2c_s}{4}-\frac{{b_v}^2}{2}+b_vc_v-c_s-2c_v+1\\
        E[\pi_M(b_s,b_v)]&=-\frac{5{b_s}^3}{12}-\frac{5{b_s}^2b_v}{2}+\frac{5{b_s}^2c_s}{8}+\frac{5{b_s}^2c_v}{4}+\frac{{b_s}^2}{8}-\frac{7b_s{b_v}^2}{2}+\frac{5b_sb_vc_s}{2}\\
        &+\frac{7b_sb_vc_v}{2}+\frac{b_sb_v}{2}-\frac{b_sc_s}{4}-\frac{b_sc_v}{2}-\frac{4{b_v}^3}{3}+\frac{7{b_v}^2c_s}{4}+2{b_v}^2c_v-\frac{{b_v}^2}{8}\\
        &-\frac{b_vc_s}{2}+\frac{b_vc_v}{4}-\frac{31c_s}{32}-\frac{31c_v}{16}+\frac{383}{384}\\
        E[\pi_M(b_s,b_v)]&=-\frac{3{b_s}^3}{4}-\frac{9{b_s}^2b_v}{2}+\frac{9{b_s}^2c_s}{8}+\frac{9{b_s}^2c_v}{4}+\frac{3{b_s}^2}{8}-\frac{15b_s{b_v}^2}{2}+\frac{9b_sb_vc_s}{2}\\
        &+\frac{15b_sb_vc_v}{2}+\frac{3b_sb_v}{2}-\frac{3b_sc_s}{4}-\frac{3b_sc_v}{2}-4{b_v}^3+\frac{15{b_v}^2c_s}{4}+6{b_v}^2c_v\\
        &+\frac{7{b_v}^2}{8}-\frac{3b_vc_s}{2}-\frac{7b_vc_v}{4}-\frac{27c_s}{32}-\frac{27c_v}{16}+\frac{125}{128}\\
        E[\pi_M(b_s,b_v)]&=-\frac{5{b_s}^3}{12}-\frac{5{b_s}^2b_v}{2}+\frac{5{b_s}^2c_s}{8}+\frac{5{b_s}^2c_v}{4}-\frac{9b_s{b_v}^2}{2}+\frac{5b_sb_vc_s}{2}+\frac{9b_sb_vc_v}{2}\\
        &-\frac{8{b_v}^3}{3}+\frac{9{b_v}^2c_s}{4}+4{b_v}^2c_v-\frac{{b_v}^2}{4}+\frac{b_vc_v}{2}-\frac{9c_s}{8}-\frac{9c_v}{4}+\frac{67}{64}\\
        E[\pi_M(b_s,b_v)]&=\frac{{b_s}^3}{4}+\frac{3{b_s}^2b_v}{2}-\frac{3{b_s}^2c_s}{8}-\frac{3{b_s}^2c_v}{4}-{b_s}^2+\frac{7b_s{b_v}^2}{2}-\frac{3b_sb_vc_s}{2}-\frac{7b_sb_vc_v}{2}\\
        &-4b_sb_v+2b_sc_s+4b_sc_v+\frac{8{b_v}^3}{3}-\frac{7{b_v}^2c_s}{4}-4{b_v}^2c_v-\frac{17{b_v}^2}{4}+4b_vc_s\\
        &+\frac{17b_vc_v}{2}-\frac{17c_s}{8}-\frac{17c_v}{4}+\frac{265}{192}\\
        E[\pi_M(b_s,b_v)]&=\frac{5{b_s}^3}{12}+\frac{5{b_s}^2b_v}{2}-\frac{5{b_s}^2c_s}{8}-\frac{5{b_s}^2c_v}{4}-\frac{5{b_s}^2}{4}+5b_s{b_v}^2-\frac{5b_sb_vc_s}{2}\\
        &-5b_sb_vc_v-5b_sb_v+\frac{5b_sc_s}{2}+5b_sc_v+\frac{10{b_v}^3}{3}-\frac{5{b_v}^2c_s}{2}-5{b_v}^2c_v\\
        &-5{b_v}^2+5b_vc_s+10b_vc_v-\frac{19c_s}{8}-\frac{19c_v}{4}+\frac{281}{192}
\end{align*}
\begin{align*}
        E[\pi_M(b_s,b_v)]&=\frac{3{b_s}^3}{4}+\frac{9{b_s}^2b_v}{2}-\frac{9{b_s}^2c_s}{8}-\frac{9{b_s}^2c_v}{4}-\frac{15{b_s}^2}{8}+9b_s{b_v}^2-\frac{9b_sb_vc_s}{2}\\
        &-9b_sb_vc_v-\frac{15b_sb_v}{2}+\frac{15b_sc_s}{4}+\frac{15b_sc_v}{2}+6{b_v}^3-\frac{9{b_v}^2c_s}{2}-9{b_v}^2c_v\\
        &-\frac{15{b_v}^2}{2}+\frac{15b_vc_s}{2}+15b_vc_v-\frac{101c_s}{32}-\frac{101c_v}{16}+\frac{229}{128}\\
        E[\pi_M(b_s,b_v)]&=\frac{5{b_s}^3}{12}+\frac{5{b_s}^2b_v}{2}-\frac{5{b_s}^2c_s}{8}-\frac{5{b_s}^2c_v}{4}-\frac{9{b_s}^2}{8}+5b_s{b_v}^2-\frac{5b_sb_vc_s}{2}\\
        &-5b_sb_vc_v-\frac{9b_sb_v}{2}+\frac{9b_sc_s}{4}+\frac{9b_sc_v}{2}+\frac{10{b_v}^3}{3}-\frac{5{b_v}^2c_s}{2}-5{b_v}^2c_v\\
        &-\frac{9{b_v}^2}{2}+\frac{9b_vc_s}{2}+9b_vc_v-\frac{65c_s}{32}-\frac{65c_v}{16}+\frac{157}{128}\\
        E[\pi_M(b_s,b_v)]&=\frac{{b_s}^3}{12}+\frac{{b_s}^2b_v}{2}-\frac{{b_s}^2c_s}{8}-\frac{{b_s}^2c_v}{4}-\frac{{b_s}^2}{4}+b_s{b_v}^2-\frac{b_sb_vc_s}{2}-b_sb_vc_v-b_sb_v\\
        &+\frac{b_sc_s}{2}+b_sc_v+\frac{2{b_v}^3}{3}-\frac{{b_v}^2c_s}{2}-{b_v}^2c_v-{b_v}^2+b_vc_s+2b_vc_v-\frac{c_s}{2}-c_v+\frac{1}{3}\\
        E[\pi_M(b_s,b_v)]&=-\frac{3{b_s}^3}{4}-\frac{9{b_s}^2b_v}{2}+\frac{9{b_s}^2c_s}{8}+\frac{9{b_s}^2c_v}{4}+\frac{3{b_s}^2}{8}-\frac{9b_s{b_v}^2}{2}\\
        &+\frac{9b_sb_vc_s}{2}+\frac{9b_sb_vc_v}{2}+\frac{3b_sb_v}{4}-\frac{3b_sc_s}{4}-\frac{3b_sc_v}{4}+\frac{9{b_v}^2c_s}{4}\\
        &-\frac{3{b_v}^2}{2}-\frac{3b_vc_s}{4}+3b_vc_v-\frac{15c_s}{16}-\frac{5c_v}{2}+\frac{17}{16}\\
        E[\pi_M(b_s,b_v)]&=\frac{{b_s}^3}{4}+\frac{3{b_s}^2b_v}{2}-\frac{3{b_s}^2c_s}{8}-\frac{3{b_s}^2c_v}{4}-{b_s}^2+\frac{9b_s{b_v}^2}{2}-\frac{3b_sb_vc_s}{2}\\
        &-\frac{9b_sb_vc_v}{2}-\frac{17b_sb_v}{4}+2b_sc_s+\frac{17b_sc_v}{4}+4{b_v}^3-\frac{9{b_v}^2c_s}{4}-6{b_v}^2c_v\\
        &-\frac{41{b_v}^2}{8}+\frac{17b_vc_s}{4}+\frac{41b_vc_v}{4}-\frac{69c_s}{32}-\frac{73c_v}{16}+\frac{181}{128}\\
        E[\pi_M(b_s,b_v)]&=\frac{3{b_s}^3}{4}+\frac{9{b_s}^2b_v}{2}-\frac{9{b_s}^2c_s}{8}-\frac{9{b_s}^2c_v}{4}-\frac{15{b_s}^2}{8}+9b_s{b_v}^2-\frac{9b_sb_vc_s}{2}\\
        &-9b_sb_vc_v-\frac{15b_sb_v}{2}+\frac{15b_sc_s}{4}+\frac{15b_sc_v}{2}+6{b_v}^3-\frac{9{b_v}^2c_s}{2}-9{b_v}^2c_v\\
        &-\frac{15{b_v}^2}{2}+\frac{15b_vc_s}{2}+15b_vc_v-\frac{101c_s}{32}-\frac{101c_v}{16}+\frac{229}{128}
        \end{align*}
\begin{align*}
        E[\pi_M(b_s,b_v)]&=\frac{5{b_s}^3}{12}+\frac{5{b_s}^2b_v}{2}-\frac{5{b_s}^2c_s}{8}-\frac{5{b_s}^2c_v}{4}-\frac{9{b_s}^2}{8}+5b_s{b_v}^2-\frac{5b_sb_vc_s}{2}\\
        &-5b_sb_vc_v-\frac{9b_sb_v}{2}+\frac{9b_sc_s}{4}+\frac{9b_sc_v}{2}+\frac{10{b_v}^3}{3}-\frac{5{b_v}^2c_s}{2}-5{b_v}^2c_v\\
        &-\frac{9{b_v}^2}{2}+\frac{9b_vc_s}{2}+9b_vc_v-\frac{65c_s}{32}-\frac{65c_v}{16}+\frac{157}{128}\\
        E[\pi_M(b_s,b_v)]&=\frac{{b_s}^3}{12}+\frac{{b_s}^2b_v}{2}-\frac{{b_s}^2c_s}{8}-\frac{{b_s}^2c_v}{4}-\frac{{b_s}^2}{4}+b_s{b_v}^2-\frac{b_sb_vc_s}{2}-b_sb_vc_v-b_sb_v\\
        &+\frac{b_sc_s}{2}+b_sc_v+\frac{2{b_v}^3}{3}-\frac{{b_v}^2c_s}{2}-{b_v}^2c_v-{b_v}^2+b_vc_s+2b_vc_v-\frac{c_s}{2}-c_v+\frac{1}{3}\\
        E[\pi_M(b_s,b_v)]&=-\frac{3{b_s}^3}{4}-\frac{9{b_s}^2b_v}{2}+\frac{9{b_s}^2c_s}{8}+\frac{9{b_s}^2c_v}{4}+\frac{3{b_s}^2}{8}-\frac{9b_s{b_v}^2}{2}\\
        &+\frac{9b_sb_vc_s}{2}+\frac{9b_sb_vc_v}{2}+\frac{3b_sb_v}{4}-\frac{3b_sc_s}{4}-\frac{3b_sc_v}{4}+\frac{9{b_v}^2c_s}{4}\\
        &-\frac{3{b_v}^2}{2}-\frac{3b_vc_s}{4}+3b_vc_v-\frac{15c_s}{16}-\frac{5c_v}{2}+\frac{17}{16}\\
        E[\pi_M(b_s,b_v)]&=\frac{{b_s}^3}{4}+\frac{3{b_s}^2b_v}{2}-\frac{3{b_s}^2c_s}{8}-\frac{3{b_s}^2c_v}{4}-{b_s}^2+\frac{9b_s{b_v}^2}{2}-\frac{3b_sb_vc_s}{2}\\
        &-\frac{9b_sb_vc_v}{2}-\frac{17b_sb_v}{4}+2b_sc_s+\frac{17b_sc_v}{4}+4{b_v}^3-\frac{9{b_v}^2c_s}{4}-6{b_v}^2c_v\\
        &-\frac{41{b_v}^2}{8}+\frac{17b_vc_s}{4}+\frac{41b_vc_v}{4}-\frac{69c_s}{32}-\frac{73c_v}{16}+\frac{181}{128}\\
        E[\pi_M(b_s,b_v)]&=-\frac{{b_s}^3}{12}-\frac{{b_s}^2b_v}{2}+\frac{{b_s}^2c_s}{8}+\frac{{b_s}^2c_v}{4}-\frac{{b_s}^2}{4}+\frac{b_s{b_v}^2}{2}+\frac{b_sb_vc_s}{2}\\
        &-\frac{b_sb_vc_v}{2}-\frac{5b_sb_v}{4}+\frac{b_sc_s}{2}+\frac{5b_sc_v}{4}+\frac{4{b_v}^3}{3}-\frac{{b_v}^2c_s}{4}-2{b_v}^2c_v\\
        &-\frac{17{b_v}^2}{8}+\frac{5b_vc_s}{4}+\frac{17b_vc_v}{4}-\frac{33c_s}{32}-\frac{37c_v}{16}+\frac{109}{128}\\
        E[\pi_M(b_s,b_v)]&=\frac{5{b_s}^3}{12}+\frac{5{b_s}^2b_v}{2}-\frac{5{b_s}^2c_s}{8}-\frac{5{b_s}^2c_v}{4}-\frac{9{b_s}^2}{8}+5b_s{b_v}^2-\frac{5b_sb_vc_s}{2}\\
        &-5b_sb_vc_v-\frac{9b_sb_v}{2}+\frac{9b_sc_s}{4}+\frac{9b_sc_v}{2}+\frac{10{b_v}^3}{3}-\frac{5{b_v}^2c_s}{2}\\
        &-5{b_v}^2c_v-\frac{9{b_v}^2}{2}+\frac{9b_vc_s}{2}+9b_vc_v-\frac{65c_s}{32}-\frac{65c_v}{16}+\frac{157}{128}
\end{align*}
\begin{align*}
        E[\pi_M(b_s,b_v)]&=\frac{{b_s}^3}{12}+\frac{{b_s}^2b_v}{2}-\frac{{b_s}^2c_s}{8}-\frac{{b_s}^2c_v}{4}-\frac{{b_s}^2}{4}+b_s{b_v}^2-\frac{b_sb_vc_s}{2}-b_sb_vc_v-b_sb_v\\
        &+\frac{b_sc_s}{2}+b_sc_v+\frac{2{b_v}^3}{3}-\frac{{b_v}^2c_s}{2}-{b_v}^2c_v-{b_v}^2+b_vc_s+2b_vc_v-\frac{c_s}{2}-c_v+\frac{1}{3}\\
        E[\pi_M(b_s,b_v)]&=-\frac{{b_s}^3}{12}-\frac{{b_s}^2b_v}{2}+\frac{{b_s}^2c_s}{8}+\frac{{b_s}^2c_v}{4}-\frac{{b_s}^2}{4}-\frac{b_s{b_v}^2}{2}+\frac{b_sb_vc_s}{2}+\frac{b_sb_vc_v}{2}\\
        &-\frac{b_sb_v}{2}+\frac{b_sc_s}{2}+\frac{b_sc_v}{2}+\frac{{b_v}^2c_s}{4}-\frac{{b_v}^2}{2}+\frac{b_vc_s}{2}+b_vc_v-\frac{3c_s}{4}-c_v+\frac{1}{2}\\
        E[\pi_M(b_s,b_v)]&=\frac{{b_s}^3}{12}+\frac{{b_s}^2b_v}{2}-\frac{{b_s}^2c_s}{8}-\frac{{b_s}^2c_v}{4}-\frac{{b_s}^2}{4}+b_s{b_v}^2-\frac{b_sb_vc_s}{2}-b_sb_vc_v-b_sb_v\\
        &+\frac{b_sc_s}{2}+b_sc_v+\frac{2{b_v}^3}{3}-\frac{{b_v}^2c_s}{2}-{b_v}^2c_v-{b_v}^2+b_vc_s+2b_vc_v-\frac{c_s}{2}-c_v+\frac{1}{3}
\end{align*}
\end{small}
\end{document}